\documentclass[fleqn,12pt,twoside]{article}
\usepackage{graphicx,mathrsfs,amsmath,amssymb}
\usepackage[headings]{espcrc1}

\title{Heavy-Quark Spectra at RHIC and Resonances in the QGP}

\author{R. Rapp\address[TAMU]{Cyclotron Institute and Physics
    Department, Texas A{\&}M University, College Station, Texas
    77843-3366, USA}, V.  Greco\address[CAT]{Laboratori Nazionali del
    Sud INFN, via S. Sofia 62, I-95123 Catania, Italy} and H. van
    Hees\addressmark[TAMU]}

\begin{document}
\maketitle

\begin{abstract}
  Thermalization and collective flow of charm ($c$) and bottom ($b$)
  quarks are evaluated from elastic parton scattering via ``$D$"- and
  ``$B$"-meson resonances in an expanding, strongly interacting
  quark-gluon plasma at RHIC. Pertinent drag and diffusion coefficients
  are implemented into a relativistic Langevin simulation to compute
  transverse-momentum spectra and azimuthal flow asymmetries ($v_2$) of
  $c$- and $b$-quarks. Upon hadronization (including coalescence and
  fragmentation) and semileptonic $D$- and $B$-decays, the resulting
  electron spectra ($R_{AA}$ and $v_2$) are compared to recent RHIC
  data.
\end{abstract}

\section{Introduction}

Among the key challenges in describing the hot and dense matter created
in Au-Au collisions at the Relativistic Heavy-Ion Collider (RHIC) is the
understanding of the microscopic interactions providing a rapid
thermalization as inferred from hydrodynamic models. Heavy quarks are
valuable probes in this respect as they are produced early in the
collision and thus sense the subsequent evolution down to rather soft
momenta.

First data on single-electron ($e^{\pm}$) spectra, associated with
semileptonic decays of $D$ and $B$-mesons, have revealed a surprisingly
large suppression (small $R_{AA}^e$) and azimuthal asymmetry
($v_2^e$)~\cite{v2-phenix,v2pre-star,jac05}.  On the one hand, within
quark coalescence models~\cite{GKR04,Mol04,Zhang05} of a hadronizing
quark plasma, the $v_2^e$ data favor the assumption that charm quarks
exhibit a degree of thermalization comparable to that of light
partons~\cite{GKR04}. On the other hand, within radiative energy-loss
calculations in a gluon plasma~\cite{Djordjevic:2004nq,arm05}, the
$R_{AA}^e$ data require significantly larger transport coefficients than
expected within perturbative Quantum Chromodynamics (pQCD).  While for
lower $p_T$ energy loss due to elastic scattering becomes parametrically
dominant (by $\sim$$1/\sqrt{\alpha}_s$)~\cite{MT04}, elastic pQCD cross
sections~\cite{Svet88,MT03} with realistic values for the strong
coupling constant ($\alpha_s$=0.3-0.5) cannot account for the observed
effects either~\cite{MT04,HGR05}. In addition, the contribution of
$B$-meson decays is expected to further reduce both suppression and
elliptic flow signals in the electron spectra.

In this talk we address the question of the microscopic interactions in
a strongly interacting Quark-Gluon Plasma (sQGP) by introducing $D$- and
$B$-meson states providing for elastic resonance cross sections for $c$-
and $b$-quarks~\cite{HR04}. Corresponding drag and diffusion
coefficients are implemented into a relativistic Langevin simulation for
semi-central Au-Au collisions at RHIC, with subsequent comparisons to
single-$e^\pm$ observables~\cite{HGR05}.

\section{Heavy-Quark Scattering in the QGP}

Lattice QCD computations of hadronic correlators suggest the survival of
mesonic resonance/bound states up to temperatures of $\sim$2$T_c$ in
both heavy- and light-quark sectors~\cite{AH-prl,KL03}, cf.~also
Refs.~\cite{SZ04,Wong04,MR05}.  Here, we simply assume the existence of
the lowest-lying, pseudoscalar $D$ ($B$) meson as a {\em resonance}
0.5~GeV above the heavy-light quark threshold~\cite{HR04}. The pertinent
effective Lagrangian with chiral and heavy-quark (HQ) symmetry then
dictates the degeneracy of the $J^P$=$0^-$ state with vector, scalar and
axial-vector partners. The 2 free model parameters are the resonance
masses ($m_{D(B)}$=2(5)~GeV, with $m_{c(b)}$=1.5(4.5)~GeV) and widths
(varied as $\Gamma$=0.4-0.75~GeV). For strange quarks we only include
pseudoscalar and vector states.  The resonant $Q$-$\bar q$ cross
sections are supplemented with leading-order pQCD scattering off
partons~\cite{com79} dominated by $t$-channel gluon exchange and
regularized by a Debye mass $m_g$=$gT$ with $\alpha_s$=$g^2/(4
\pi)$=0.4. When evaluating drag and diffusion coefficients in a
Fokker-Planck approach~\cite{Svet88}, the resonances 
reduce HQ thermalization times by a factor of $\sim$3 below pQCD
scattering~\cite{HR04}.

\begin{figure}[!tb]
\begin{minipage}{0.47\textwidth}
\includegraphics[width=\textwidth]{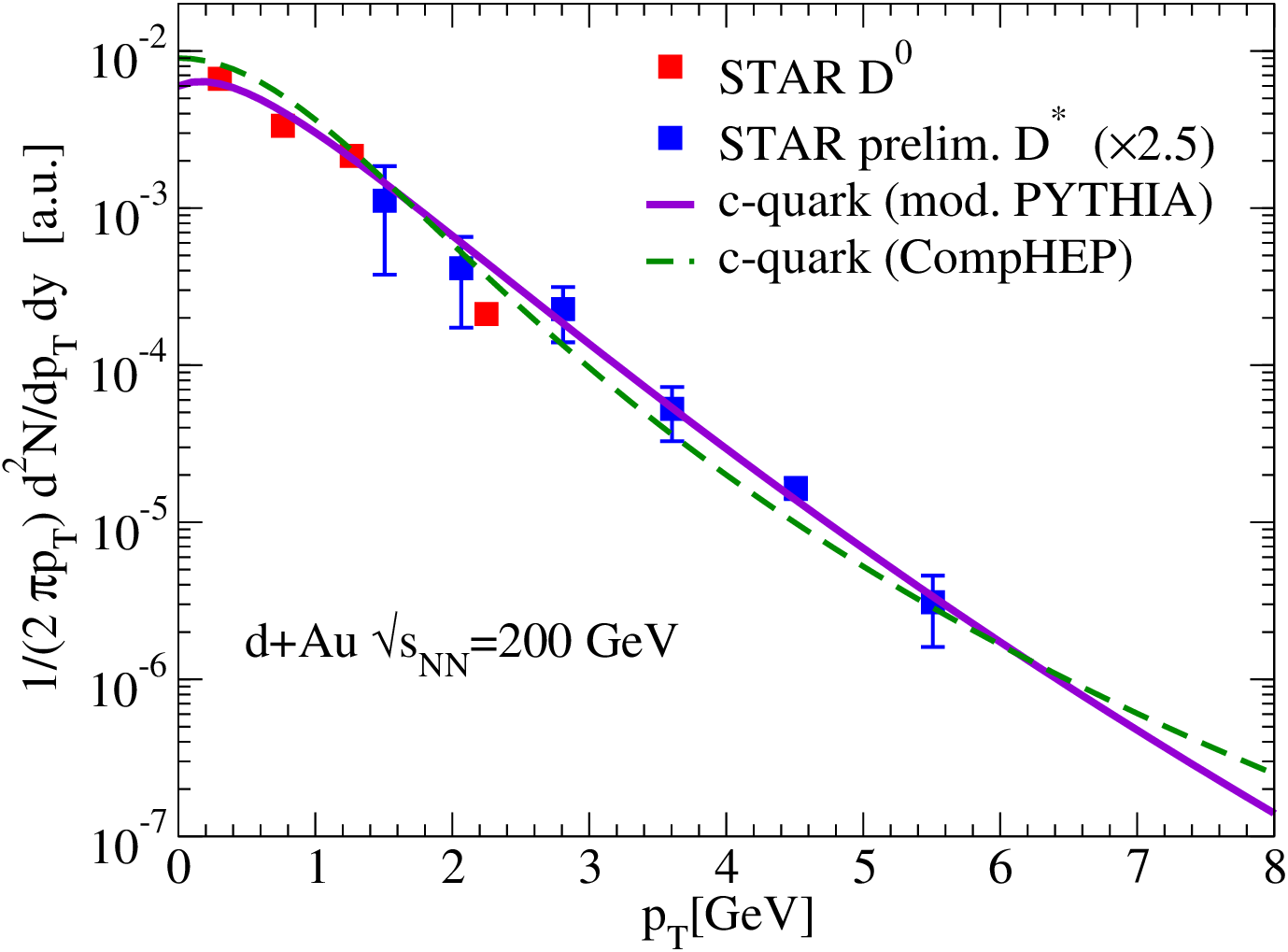}
\end{minipage}\hfill
\begin{minipage}{0.48\textwidth}
\includegraphics[width=\textwidth]{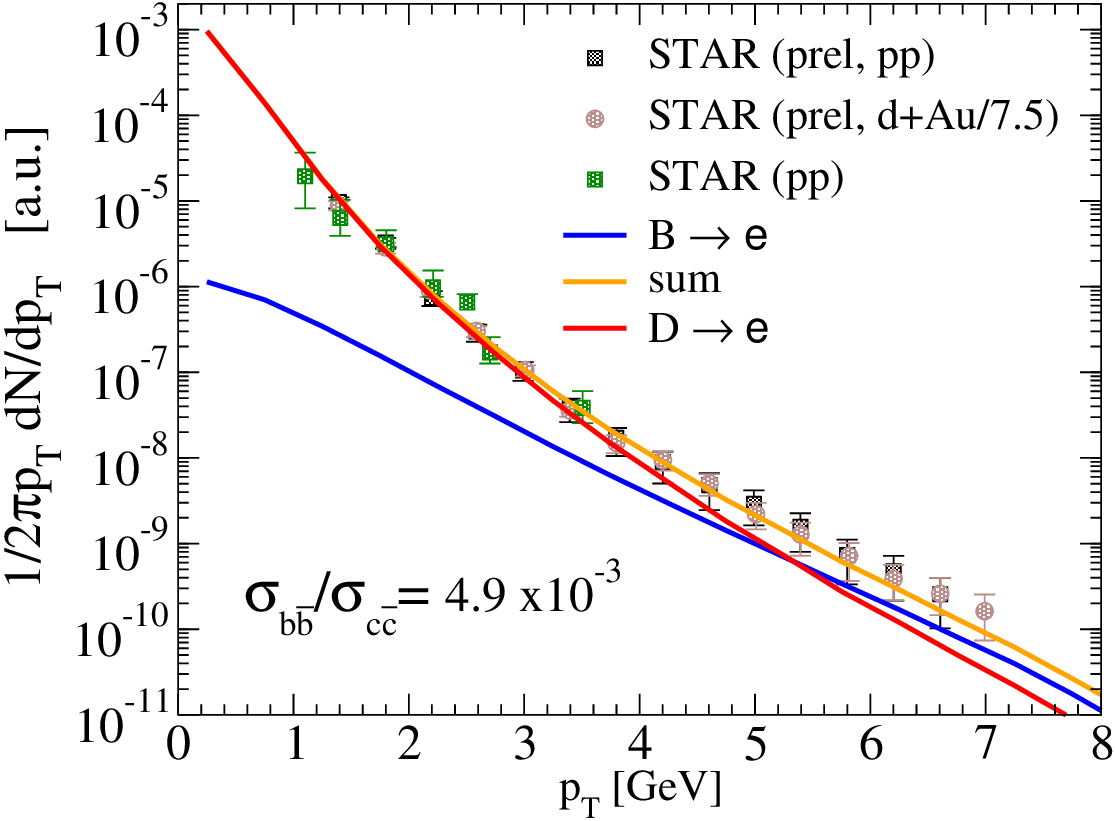}
\end{minipage}
\caption{Our fit of initial $c$- and $b$-quark spectra using experimental
  spectra for $D$- and $D^*$-mesons~\cite{star-D,star-dAu} (left panel) and
  single-$e^{\pm}$~\cite{star-dAu-pp,star-dAu} (right panel).}
\label{fig.1}
\end{figure}
The heavy quarks are implemented into $b$=7~fm Au-Au
collisions at RHIC via relativistic Langevin simulations~\cite{MT04} in
an isentropically expanding isotropic QGP fireball. The expansion
parameters are fixed to closely resemble the time evolution of radial
and elliptic flow in hydrodynamic models~\cite{kol00}, with an
appropriate hadron multiplicity at chemical freezeout. A formation time
of 1/3~fm/c translates into an initial temperature of $T_0$=340~MeV,
based on an ideal QGP equation of state with 2.5 flavors. The Langevin
process is simulated in the H{\"a}nggi-Klimontovich
realization~\cite{hae05}, i.e., in the local rest frame of the bulk
matter with an update of HQ position and momentum given by
\begin{equation}
\delta \vec{x}=\frac{\vec{p}}{E} \, \delta t, \quad \delta
\vec{p}=-A(t,\vec{p}+\delta \vec{p}) \, \vec{p} \, \delta t+\delta
\vec{W}(t,\vec{p}+\delta \vec{p}) \ 
\end{equation}
($E$: HQ energy); $\delta \vec{W}$ is a random force distributed
according to Gaussian noise,
\begin{equation}
P(\delta \vec{W}) \propto \exp \left [-\hat{B}_{jk} \delta W^j
    \delta W^k/(4 \delta t) \right] \ .
\end{equation}
The drag coefficient (inverse relaxation time), $A$, and the
diffusion-coefficient matrix,
\begin{equation}
B_{jk}=(\hat{B}^{-1})_{jk}=B_0(\delta^{jk}-\hat{p}^j \hat{p}^k)+B_1
\hat{p}^j \hat{p}^k,
\end{equation}
follow from the microscopic model sketched above~\cite{HR04}.  The
thermal equilibrium limit is enforced by setting the longitudinal
diffusion coefficient to $B_1$=$T E A$~\cite{MT04}.  HQ momenta are
Lorentz-boosted to the laboratory frame according to the local bulk
matter velocity.

To determine the initial HQ transverse-momentum ($p_T$) 
distributions and the relative
magnitude of the $c$- and $b$-quark spectra, we use a modified PYTHIA
$c$-quark spectrum and $\delta$-function fragmentation to fit STAR $D$
and $D^*$ spectra in d-Au collisions~\cite{star-D,star-dAu} (left panel of
Fig.~\ref{fig.1}). The pertinent $e^{\pm}$-decay spectra saturate data
from $p$-$p$ and d-Au up to $p_T^e \simeq
3.5$~GeV~\cite{star-dAu-pp,star-dAu} (right panel of Fig.~\ref{fig.1}).
The missing yield at higher $p_T$ is then attributed to $B$-meson
decays, resulting in a cross section ratio of $\sigma_{b
  \bar{b}}/\sigma_{c\bar{c}}$=$4.9 \cdot 10^{-3}$ and implying a
crossing of $D$- and $B$-decay electrons at $p_T$$\simeq$5~GeV.

\section{Hadronization and Single-Electron Spectra}
\begin{figure}[!tb]
\begin{minipage}{0.48 \textwidth}
\includegraphics[width=\textwidth]{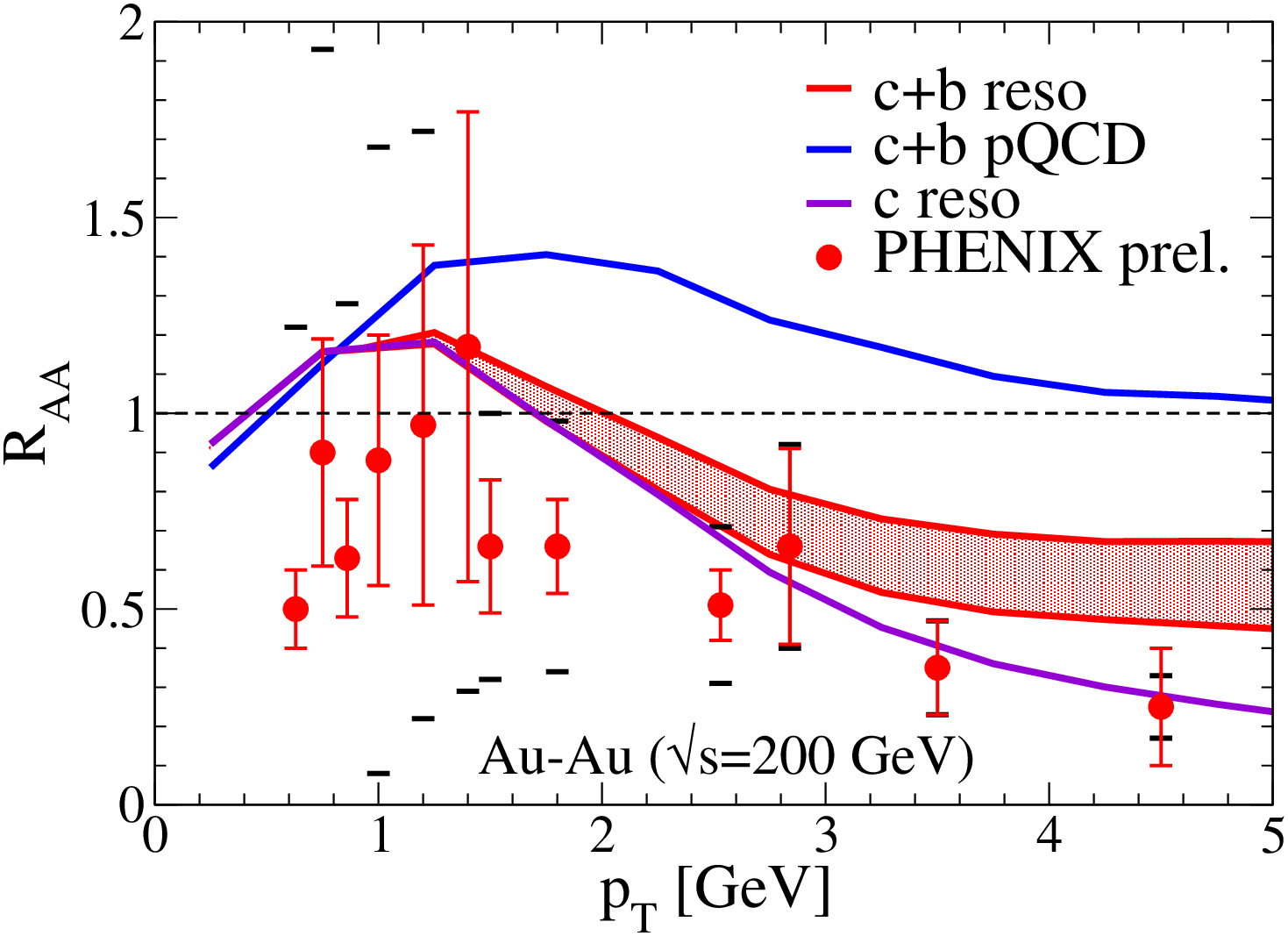}
\end{minipage}\hfill
\begin{minipage}{0.465 \textwidth}
\includegraphics[width=\textwidth]{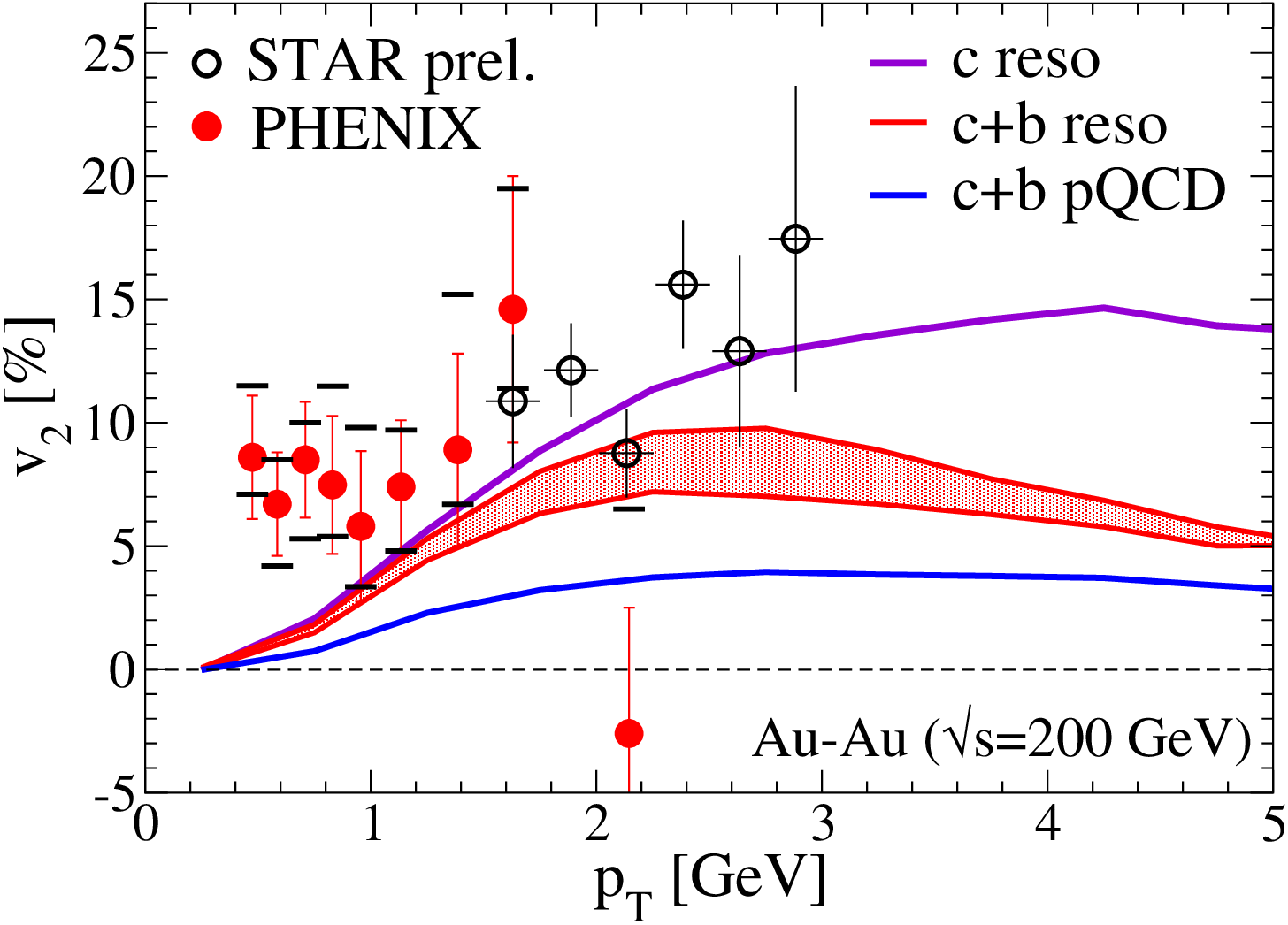}
\end{minipage}
\caption{Nuclear modification factor ($R_{AA}$, left panel) and 
  elliptic flow ($v_2$, right panel) of semileptonic $D$- and $B$-meson
  decay electrons in $b$=7~fm $\sqrt{s_{NN}}$=200~GeV Au-Au
  collisions assuming different elastic heavy-quark interactions in the
  QGP with subsequent coalescence+fragmentation hadronization, compared
  to PHENIX and STAR data~\cite{v2-phenix,v2pre-star,jac05}.}
\label{fig.2}
\end{figure}
To compare our results for the final HQ $p_T$-spectra and $v_2$,
recorded at the end of the mixed phase, to measured $e^{\pm}$ spectra in
Au-Au collisions, we hadronize $b$- and $c$-quarks using the coalescence
model of Ref.~\cite{GKR04} based on Ref.~\cite{Greco:2003mm} for the
light-quark distributions. While this essentially exhausts the HQ yields
at low $p_T$, the decreasing light-quark phase-space density at higher
$p_T$ implies unpaired $c$- and $b$-quarks which we hadronize via
$\delta$-function fragmentation. The single-$e^{\pm}$ spectra follow
from $D$- and $B$-meson 3-body decays, cf.~Fig.~\ref{fig.2}.  Compared
to elastic pQCD rescattering alone, resonance effects manifest
themselves as a substantial increase (decrease) in $v_2^e$ ($R_{AA}^e)$,
while coalescence further amplifies $v_2^e$ but also increases
$R_{AA}^e$.  The bottom contributions reduce the effects starting from
electron momenta of about $\sim$3~GeV. Overall, the main trends in the
data are reasonably well reproduced.

\section{Conclusions}
Based on elastic resonant interactions in the sQGP we have evaluated
$c$- and $b$-spectra in an expanding fireball at RHIC employing
relativistic Langevin simulations. The much increased cross sections
compared to pQCD lead to $c$-quark $R_{AA}$ and $v_2$ of down to 0.2 and
up to 10\%, respectively, while $b$-quarks are less affected.  After
subsequent hadronization in a combined quark-coalescence and
fragmentation scheme followed by semileptonic decays, we have found that
resonant interactions may play an important role in the simultaneous
understanding of the nulcear modification factor and elliptic flow of
heavy-quark observables (including single electrons) at RHIC, and thus
open a promising window on the microscopic properties of the sQGP
including its rapid thermalization behavior.
\\

{\bf Acknowledgments.}
One of us (HvH) has been supported in part by a F.-Lynen Fellowship
of the A.-v.-Humoboldt Foundation.
This work has been supported in part by a U.S. National Science
Foundation CAREER award under grant PHY-0449489.


\begin{thebibliography}{99}

\bibitem{v2-phenix}
S.S.~Adler {\it et al.} [PHENIX Collaboration], Phys.\ Rev.\ C {\bf 72},
024901 (2005).

\bibitem{v2pre-star}
F.~Laue {\it et al.} [STAR Collaboration], J.\ Phys.\ G {\bf 31}, S27 (2005).

\bibitem{jac05}
B.~Jacak {\it et al.} [PHENIX Collaboration], to appear in Proc. of 5$^{\mathrm{th}}$
International Conference on Physics and Astrophysics of Quark Gluon
Plasma (2005), nucl-ex/0508036.


\bibitem{GKR04}
V.~Greco, C.M.~Ko and R.~Rapp, Phys. Lett. \textbf{B595}, 202 (2004).

\bibitem{Mol04}
D.~Molnar, J. Phys. \textbf{G31}, S421 (2005).

\bibitem{Zhang05}
B.~Zhang, L.W.~Chen and C.M.~Ko, Phys. Rev. C \textbf{72}, 024906 (2005).


\bibitem{Djordjevic:2004nq}
 M.~Djordjevic, M.~Gyulassy and S.~Wicks,
 Phys. Rev. Lett.  {\bf 94}, 112301 (2005).

\bibitem{arm05}
N.~Armesto {\it et al.},
Phys. Rev. D \textbf{71}, 054027 (2005).

\bibitem{MT04}
G.D.~Moore and D.~Teaney, Phys. Rev. D \textbf{71}, 064904 (2005)

\bibitem{Svet88}
B.~Svetitsky, Phys. Rev. D \textbf{37}, 2484 (1988).

\bibitem{MT03}
M.G.~Mustafa and M.H.~Thoma, Acta Phys. Hung. A \textbf{22}, 93 (2005).

\bibitem{HGR05}
H.~van Hees, V.~Greco and R.~Rapp (2005), nucl-th/0508055.

\bibitem{HR04}
H.~van Hees and R.~Rapp, Phys. Rev. C \textbf{71}, 034907 (2005).

\bibitem{star-D} 
J.~Adams {\it et al.}  [STAR Collaboration], Phys. Rev.
Lett. {\bf 94}, 062301 (2005).

\bibitem{star-dAu}
A.~Tai {\it et al.}  [STAR Collaboration], J. Phys. G {\bf 30}, S809 (2004).

\bibitem{star-dAu-pp}
A.A.P. Suaide {\it et al.} [STAR Collaboration] J. Phys. G \textbf{30},
S1179 (2004).


\bibitem{AH-prl}
M.~Asakawa and T.~Hatsuda, Phys. Rev. Lett. \textbf{92}, 012001 (2004).

\bibitem{KL03} 
F.~Karsch and E.~Laermann, hep-lat/0305025.

\bibitem{SZ04}
E.V.~Shuryak and I.~Zahed, Phys. Rev. C \textbf{70}, 021901(R) (2004).

\bibitem{Wong04}  
C.Y.~Wong (2004), hep-ph/0408020.

\bibitem{MR05}
M.~Mannarelli and R.~Rapp (2005), hep-ph/0505080.

\bibitem{com79}
B.L.~Combridge, Nucl. Phys. \textbf{B151}, 429 (1979).

\bibitem{kol00}
P.F.~Kolb, J.~Sollfrank and U.~Heinz, Phys. Rev. C \textbf{62}, 054909
(2000).

\bibitem{hae05}
J.~Dunkel and P.~H{\"a}nggi, Phys. Rev. E \textbf{71}, 016124 (2005).

\bibitem{Greco:2003mm}
V.~Greco, C.~M.~Ko and P.~Levai, Phys. Rev. C {\bf 68}, 034904 (2003).


\end{thebibliography}
\end{document}